# A greener process for poly-L-lactic acid production and chemical upcycling under mild conditions using highly active alkali-metal based catalysts

*Christian Rentero,[†] Licia Gaeta,[†,‡] Miguel Palenzuela, [†, §] Valentina Sessini*[†] and Marta E.G. Mosquera*[†]*

[†] Departamento de Química Orgánica y Química Inorgánica, Instituto de Investigación en Química "Andrés M. del Río" (IQAR), Universidad de Alcalá, Campus Universitario, Alcalá de Henares, Madrid 28871, Spain. Email: martaeg.mosquera@uah.es, valentina.sessini@uah.es

[‡] Department of Chemistry and Biology, "A. Zambelli", University of Salerno, via Giovanni Paolo II, 132, Fisciano, SA 84084, Italy - Department of Industrial Engineering, University of Padova, via Gradenigo, 6/a, Padova, PD 35131, Italy.

[§] SUSPOL. Polímeros Sostenibles S.L. Calle Siete Esquinas, Alcalá de Henares, Madrid 28801, Spain

**Abstract**

Non-toxic potassium and sodium metal compounds have been prepared with a bulky aryloxide ligand MOR (**1**, **2**), as well, their counterparts containing a crown ether bonded to the alkali metal (**3**, **4**), and fully characterized. The activity of compounds **1-4** as catalysts for the ring-opening polymerization (ROP) process of LLA have been studied, showing that they are extremely active and able to polymerize LLA within a few minutes under mild conditions, achieving PLLA with high molecular weight, similar to the commercial ones. For derivatives **3** and **4**, the crown ether coordination to the alkali metal affects the nuclearity of the compounds and consequently its activity, giving a more controlled polymerization. As well the use of BnOH as co-initiator allowed a better control over the polymerization. Detailed studies of the polymerization mechanism have been performed, which confirmed an anionic mechanism in absence of a co-initiator. Furthermore, the nature of the mechanisms provokes the epimerization of the lactide and the existence of *D*-isomers in the PLLA. Since the percentage of *D*-isomer content in the polymeric chain as well as its distribution can strongly change the properties of PLLA a detailed analysis has been performed. As expected, the different isotacticity leads to a strong variation on the thermal properties. We have also compared the mechanical properties of the different synthetized PLLA, and we clearly observed that the epimerization reactions lead to an increase of the PLLA flexibility compared with commercial PLAs. Furthermore, the alkali metal compounds prepared not only polymerize, as well in the presence of an alcohol they can depolymerized PLLA within 15 minutes to give alkyl lactates, allowing the easy upcycling of commercial PLLAs.

**Introduction**

Traditional plastics, renowned for their versatility, durability, and cost-effectiveness, have been globally manufactured since the early 20th century. However, their prolonged durability has led to an exponential rise in plastic waste. Typically derived from finite fossil fuels, these non-biodegradable plastics contribute to the formation of persistent microplastics, amongst other environmental issues [1]. In this context, bioplastics arise as a more environmentally friendly alternative. They can be derived from renewable feedstock and can be biodegradable [2]. However, these materials come with some drawbacks, including generally higher manufacturing costs compared to traditional plastics [3]. Hence the search for a more efficient synthesis is an open challenge. In this context, catalysis enables a very effective pathway to enhance production efficiency [4].

Polylactic acid (PLA) stands out as one of the most popular bioplastics due to its biocompatibility and biodegradability. These inherent properties make it suitable for applications in packaging, single-use utensils, and medical implants [5]. Achieving precise control during its synthesis is essential to obtain a well-defined structure, which leads to sustainable and functional materials with advanced properties [6]. PLA is typically produced through the Ring-Opening Polymerization (ROP) of lactide, a cyclic diester derived from lactic acid. Metal-based catalysts are frequently employed for the ROP of lactide, with stannous octanoate, [$Sn(Oct)_2$], being the most commonly used catalyst at an industrial scale. However, this metal is known to have associated toxicity and can present potential health risks during the handling and manufacturing of PLA [7]. Hence, there is a clear interest in developing novel compounds based on less toxic metals for the effective lactide polymerization. Different examples of efficient catalytic systems utilizing metals such as zinc, magnesium, or alkali metals have been reported [8-12]. Considering the potential use of PLA in biomedical applications, the selection of a biocompatible metal is crucial, with sodium and potassium complexes emerging as promising choices.

Most studies focused on the polymerization of *rac*-lactide (*rac*-LA), the equimolar mixture of *L*- and *D*-isomers [8-9]. Nevertheless, *L*-lactide (LLA) is also cost-effective [13], and there are several examples of alkali-based complexes that show high activities for LLA polymerization. However, achieving high molecular weight poly-*L*-lactic acid (PLLA) remains a challenge using this type of catalyst [8, 10, 14]. On the other hand, the stereoregularity and microstructure are key points of the industrial production of PLLA. In fact, the percentage of *D*-isomer content in the polymeric chain as well as its distribution can strongly change the properties of PLLA [15-16]. As such, its mechanical and thermal properties, crystallinity and even its degradation characteristics are highly dependent on the *D*-content. Thus, controlling this content, highly isotactic semi-crystalline or heterotactic amorphous PLLA can be produced [17-18].

Once the PLA-based products reach the end of their life cycle, managing this bioplastic becomes a crucial step when following a circular economy model. Although PLA is considered biodegradable, specific conditions are required, including optimal temperature, humidity, and microbial activity, which are only available in industrial composting facilities. Consequently, PLA does not readily degrade in ordinary environmental conditions, including marine or home compost conditions [3]. As an alternative for managing PLA waste, chemical recycling offers an efficient method to recover the raw material, enabling the production of new materials with identical properties [19]. Nevertheless, the depolymerization of PLA is challenging and requires drastic conditions [20]. Consequently, various upcycling strategies can be employed to transform the polymer into different value-added products through chemical reactions such as alcoholysis, hydrolysis, or hydrogenation [21-23]. Specifically, catalytic alcoholysis of PLLA with methanol results in the production of methyl-*L*-lactate (Me-LLA). This added-value product finds diverse applications across various fields, serving as cooling or flavoring agent in food, a fragrance in cosmetics, a green solvent for cellulosic materials and for organic synthesis, including being a platform for the 1,2-propanediol production, amongst others [24]. Most of the reported catalysts for PLA methanolysis are based on zinc and magnesium, and few of them can fully transform PLLA into methyl lactate. However, almost all the complexes reported require high temperatures, extended reaction times, and/or a large excess of methanol as a reagent to achieve high conversion and selectivity. Additionally, ethyl-*L*-lactate (Et-LLA) can also be obtained through the alcoholysis of PLLA with ethanol, an added-value product that has also been reported as a solvent alternative for several applications [25].

Aiming to design and develop a process to produce PLLA and its chemical upcycling using alkali-based complexes under mild conditions, in this work we present readily available sodium and potassium phenoxide compounds that are extremely active for the ROP of LLA to give high molecular weight PLLA, suitable for potential biomedical applications. As well, we explored the impact of incorporating crown ethers into the metal coordination sphere and benzyl alcohol, BnOH, as co-initiator to achieve more controlled polymerizations, leading to polymers with narrow dispersity values. Besides, by tuning the experimental conditions it is possible to control the *D*-isomer content in the final polymer and hence the thermal and mechanical properties, as shown in the detailed studies performed in this work. Furthermore, this catalytic system resulted also effective for the chemical upcycling of PLLA, capable of depolymerizing PLLA via alcoholysis to generate added-value products, depending on the alcohol used. Our catalytic system showed high activity, enabling upcycling at lower temperature, with reduced reaction times, and lower alcohol ratio, in comparison to the conditions previously reported in the literature.

**Materials and Methods**

*Materials*

All manipulations were conducted under inert atmosphere using standard Schlenk line techniques ($O_2 < 3$ ppm) in conjunction to a MBraunUnilab-MB-20-G glove box ($O_2 < 0.6$ ppm). All solvents were rigorously dried prior to use using an MBraun Solvent Purification System. All the reagents employed in the synthesis of the preligand, and the alkali-based compounds were commercially obtained and used without further purification. *L*-lactide (>98%) was purchased from TCI Chemicals, and was purified by recrystallization in toluene and sublimation at 100 °C. The purified monomer was stored in the glovebox. Poly-*L*-lactic acid (Luminy L105, $M_w \approx 80$ kDa, *D*-isomer content <1%) was supplied by Total Corbion PLA DV (Gorinchem, Netherlands). Methanol (MeOH), ethanol (EtOH) and benzyl alcohol (BnOH) were dried with $CaH_2$, distilled under vacuum, and stored over 3 Å molecular sieves.

*Characterization techniques*

*Nuclear Magnetic Resonance (NMR)* spectra were recorded on a Bruker 400 Ultrashield ($^1$H 400 MHz, $^{13}$C 101 MHz) at room temperature. Chemical shifts (γ) are reported in ppm relative to tetramethylsilane (TMS) and were referenced internally using $C_6D_6$, $CDCl_3$, DMSO-$d_6$ as solvents.

*Single-Crystal X-ray Structure Determination (XRD)*. Data collection was performed at 200(2) K, with the crystals covered with per-fluorinated ether oil. Single crystals of the corresponding compound were mounted on a Bruker D8 Venture single crystal diffractometer equipped with a Mo-Kα radiation (λ = 0.71073 Å). Multiscan absorption correction procedures were applied to the data [26]. The structure was solved using the WINGX package [27], by direct methods (SHELXS-13) and refined using full-matrix least-squares against $F^2$ (SHELXL-16) [28]. For both compounds the data quality was poor as the crystal only diffracted to low theta values (18°), in both cases two independent molecules appear in the asymmetric unit. All non-hydrogen atoms were anisotropically refined, including several disordered carbon atoms. Hydrogen atoms were geometrically placed and left riding on their parent atoms Both compounds crystallized with disordered solvent molecules (Et$_2$O for **2** and 2 molecules of hexane for **4**) that could not be modelled so the Squeeze procedure was applied to remove their contributions from the structure factors [29]. Besides for **2** a disorder ether molecule is coordinated to the potassium atom; the positional disorder of this molecule was not treated. Full-matrix least-squares refinements were carried out by minimizing $\sum w(Fo^2 - Fc^2)^2$ with the SHELXL-16 weighting scheme and stopped at shift/err < 0.001. The final residual electron density maps showed no remarkable features. Crystallographic data for the structure reported in this paper has been deposited with the Cambridge Crystallographic Data Centre as supplementary publication numbers CCDC: 2369286 (**4.2hexane**) and 2369287 (**2.Et$_2$O**)

*Size-exclusion chromatography (SEC)* The molecular weights ($M_n$ and $M_w$) and dispersity values (Đ = $M_w/M_n$) of the polymers were determined by SEC on an Agilent 1260 Infinity II high-speed liquid chromatograph LC System equipped with two gradient columns PL gel 5 μm (Mixed-D and Mixed-C) connected in series. Tetrahydrofuran (THF) was the eluent, and the samples solutions were injected with a 1 mL min$^{-1}$ flow rate at 30 °C. The calibration was performed using polystyrene (PS) standards. To check the suitability of the method, commercial PLLA Luminy L105 was checked. The values obtained (Mn= 67.5 kDa, Mw = 109.9 kDa, Đ = 1.63) are similar to the ones given in the specifications.

*Mass-assisted laser desorption/ionization time-of-flight (MALDI-TOF)* mass spectra were obtained on a Bruker ULTRAFLEX III TOF/TOF, with an NdY AG laser using DCTB as matrix and NaI or KI as cationization agent, depending on the metal catalysts employed in the polymerization.

*Specific optical rotation* was determined using a polarimeter Perkin Elmer (341) in chloroform (0.010 g of PLLA in 5 mL of solvent) at 25 ˚C. Equation (1) was used to determine the rate of isomer *D*.

$$D\% = 100 \times \left[\frac{([\alpha]_D^{25})_{PLLA} - ([\alpha]_D^{25})_{PLLA_{synt}}}{2\,([\alpha]_D^{25})_{PLLA}}\right] \qquad \text{Equation (1)}$$

Where the specific rotation of pure PLLA is considered as $([\alpha]_D^{25})_{PLLA} = -156°$ [30].

*Thermogravimetric analyses (TGA)* were carried out on a TGA55 analyzer (TA Instrument). The dynamic experiments were performed using about 10 mg of sample from 25 to 600 ˚C at 10 ˚C min$^{-1}$ under nitrogen atmosphere (60 mL min$^{-1}$). The degradation temperatures were determined at a 5% weight loss ($T_{5\%}$) and at the maximum degradation rate ($T_{max}$) from the first derivative of the TGA curves (DTG).

*Differential Scanning Calorimetry (DSC)* analyses were carried out on a TA Instrument Discovery DSC25, performing heating/cooling/heating cycles program in the range of 0 to 200 ˚C with a heating/cooling rate of 10 ˚C min$^{-1}$ and run under nitrogen purge (50 mL min$^{-1}$). Glass transition, cold crystallization and melting temperatures ($T_g$, $T_{cc}$ and $T_m$, respectively) were measured from the second heating scan of the thermograms. The degree of crystallinity ($\chi_c$) was determined by using Equation (2):

$$\chi_c = 100 \times \left[\frac{\Delta H_m - \Delta H_{cc}}{\Delta H_m^{100}}\right] \qquad \text{Equation (2)}$$

where $\Delta H_m$ is the enthalpy of fusion, $\Delta H_{cc}$ is the enthalpy of cool crystallization, $\Delta H_m^{100}$ is the enthalpy of fusion of a 100 % crystalline PLA, taken as 93 J g$^{-1}$ [31].

*Mechanical properties* were determined by tensile tests conducted at room temperature. The specimens were characterized by an Instron 3345 universal testing machine equipped with a 100 N load cell. The crosshead displacement rate was set at 10 mm min$^{-1}$, with an initial length between clamps of 12 mm. The mechanical properties reported are normalized to the sample dimensions (length, width and thickness) following standard ISO 37: dumbbell Type 3. Young's modulus (E, slope of the curve until 2% deformation), yield strength ($\sigma_y$) and elongation ($\varepsilon_y$), as well as ultimate strength ($\sigma_b$) and elongation at break ($\varepsilon_b$) as average values of at least four determinations of each sample.

*Scanning electron microscopy (SEM).* The electrospun fibers previously sputter-coated with gold (Polaron SEM coating system, 1.4 kV - 1.8 mA - 120 min, thickness ≈ 500 Å) were studied by SEM (Jeol JSM-IT500 instrument) operating at 10 kV.

**Results and Discussion**

*Synthesis of the compounds*

The bulky preligand 2,6-bis(diphenylmethyl)-4-tert-butylphenol, HOAr, was synthesized as previously reported [32]. Subsequently, the sodium, [NaOAr] (**1**), and the potassium, [KOAr] (**2**), complexes were obtained by treating the preligand with the corresponding alkali metal hydrides at low temperature [32-33]. The deprotonation of HOAr resulted in the coordination of the phenoxide ligand to the corresponding metal center. Besides, after the formation of **1** and **2**, the derivatives **3** and **4** were prepared by the addition *in situ* of 15-crown-5 and 18-crown-6 ethers respectively in the reaction media (Scheme 1).

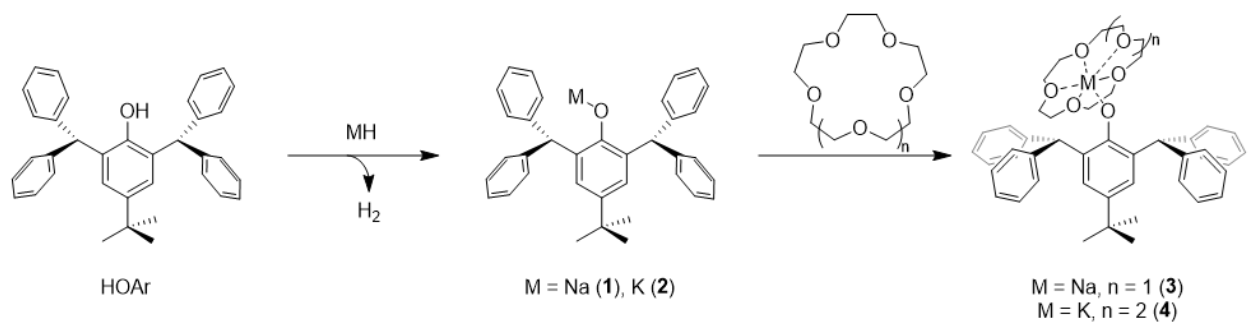

Scheme 1. Synthesis of the alkali metal-based compounds.

The formation of each complex was confirmed by $^1$H NMR by the disappearance of the signal corresponding to the hydroxyl group of HOAr, and the shift of the singlets corresponding to the protons of the *tert*-butyl and methine groups to higher and lower field, respectively (Figures S1-S4). For compounds **3** and **4**, two singlets are clearly observed at 2.95 and 3.05 ppm, respectively, that corresponds to the crown ethers. Furthermore, the coordination of the crown ether was confirmed by DOSY experiments where all the signals for compounds **3** and **4** diffuse at the same value including the ones for the crown ether (Figures S5-S11).

Although the structure in solid state of **1** has been previously reported [32], appropriate crystals of **2** and **4** were isolated from a concentrated solution to determine their structures in the solid state by single-crystal X-Ray Diffraction. Though the crystal structures do not have a good quality due to the poor diffraction of the crystals, the connectivity of the atoms is clearly determined. In both cases two molecules appear in the asymmetric unit, for compound **2**, a dinuclear derivative is observed where the bulky phenoxide ligands act as bridging ligands (Figure 1). Besides, a disordered ether molecule is also coordinated to the potassium metal. The alkali metal is further stabilized by π interactions with two phenyl groups from the ligand substituents, as has been previously observed for this ligand.[33] In fact, the two phenyl rings place themselves in a kind of pocket where the alkali metal fits with a symmetric coordination. Compound **4** is mononuclear and the coordination of the potassium atom to the crown ether is evident. Since the metal is trapped by the crown ether, the interaction with the phenyl groups in the *ortho* position is hampered (Figure S44).

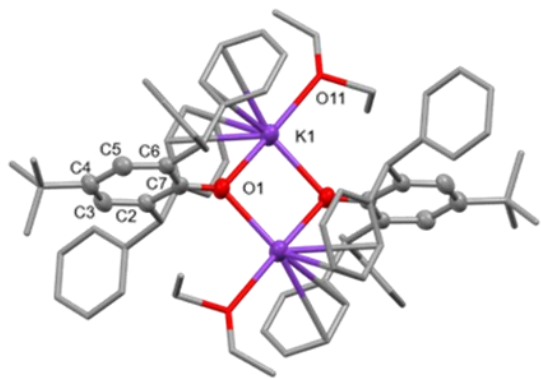

Figure 1. ORTEP plot of **2** showing thermal ellipsoid plots (30% probability). Hydrogen atoms have been omitted, the substituents of the phenoxide ligand and the coordinated ether molecule are depicted as sticks for clarity.

*Polymerization studies*

The initial studies involved LLA polymerization in the absence of co-initiator, using alkali-based complexes **1** and **2** as catalysts. As shown in Table 1, both catalysts displayed high activity, achieving full conversion in a few minutes. Furthermore, both compounds demonstrated a similar behavior, leading to unexpectedly high number average molecular weights ($M_{n,exp.}$) of the polymers that surpassed the theoretical values ($M_{n,theo}$). Moreover, as the monomer-to-catalyst ratio increased, reaction times extended, and lower dispersity values were observed. Although the process is not perfectly controlled, there is a correlation between the increasing monomer ratio and the $M_n$. However, there is a limit to this ratio, beyond which full conversion cannot be achieved.

When comparing their performances, **1** exhibited higher activity than **2** as the reaction is completed faster and can operate at a higher monomer-to-catalyst ratio, enabling the production of polymers with higher $M_n$. Moreover, when using **1** as catalyst, for the full conversion reactions a better correlation is observed between the molecular weight increase and the polymerization degree.

Table 1. ROP of LLA using alkali-based complexes as catalysts. [a]

| Ent. | Cat. | [LLA]:[Cat] | t (min) | conv.[b] (%) | $M_{n, theo}$ [c] (kDa) | $M_{n, exp}$ [d] (kDa) | Đ [d] |
|---|---|---|---|---|---|---|---|
| 1 | 1 | 100:1 | 0.5 | >99 | 14.4 | 70.5 | 1.91 |
| 2 | 1 | 200:1 | 1 | >99 | 28.8 | 99.0 | 1.76 |
| 3 | 1 | 400:1 | 2 | >99 | 57.6 | 165.4 | 1.75 |
| 4 | 1 | 800:1 | 8 | 76 | 87.6 | 86.8 | 1.71 |
| 5 | 2 | 100:1 | 1.5 | >99 | 14.4 | 57.5 | 2.03 |
| 6 | 2 | 200:1 | 2 | >99 | 28.8 | 63.9 | 1.87 |
| 7 | 2 | 400:1 | 4 | 88 | 50.5 | 81.5 | 1.73 |

[a] All polymerizations were performed in $CH_2Cl_2$ at room temperature, [LLA] = 2 M.
[b] Determined by $^1H$ NMR spectroscopy. [c] $M_{n, theo}$ = ([LLA]$_0$/[Cat]$_0$) · $MW_{,LLA}$ · %Conv.
[d] Determined by GPC in THF, using polystyrene standards.

The strategy of incorporating a crown ether into the coordination sphere has been reported to provide a constrained environment around the alkali metal, which enables better control over the polymerization process [9, 14, 34]. Aiming to achieve more control over the polymerization we explored the performance of the compounds, **3** and **4**. Although these compounds are able to polymerize lactide very fast, they exhibited lower activity compared to their counterparts **1** and **2**, requiring longer reaction times, and full conversion could not always be reached (Table 2). However, better control was achieved, allowing the synthesis of polymers with lower dispersities and molecular weights closer to the calculated values. Once again, the sodium compound exhibited higher activity.

Table 2. LLA polymerization employing crown ether derivatives.[a]

| Ent. | Cat | [LLA]:[Cat] | t (min) | conv.[b] (%) | $M_{n, theo}$[c] (kDa) | $M_{n, exp}$[d] (kDa) | Đ[d] |
|---|---|---|---|---|---|---|---|
| 1 | 3 | 100:1 | 4 | >99 | 14.4 | 17.2 | 1.58 |
| 2 | 3 | 200:1 | 6 | 95 | 28.8 | 22.8 | 1.46 |
| 3 | 4 | 100:1 | 8 | 83 | 12.0 | 12.0 | 1.48 |
| 4 | 4 | 200:1 | 16 | 70 | 20.1 | 13.7 | 1.49 |

[a] All polymerizations were performed in $CH_2Cl_2$ at room temperature, [LLA] = 2 M.
[b] Determined by $^1H$ NMR spectroscopy. [c] $M_{n, theo}$ = ([LLA]$_0$/[Cat]$_0$) · $MW_{,LLA}$ · %Conv.
[d] Determined by GPC in THF, using polystyrene standards.

Another alternative to enhance the control over polymerization is the use of a co-initiator. Therefore, we explored the influence of BnOH, a widely used co-initiator for the ROP. In our previous studies, we observed that the addition of just one equivalent leads to the presence of two competing mechanisms: an anionic mechanism or an activated-monomer mechanism [14]. To ensure that the polymerization proceeded via an activated-monomer mechanism, we added two equivalents of BnOH per equivalent of catalyst. The polymerizations showed in Table 3 were conducted using **1** and **2** as catalysts and BnOH as a co-initiator. Although similar conversions and reaction times were observed compared to the results showed in Table 1, the molecular weights and dispersity values indicated that the process was more controlled, particularly in the full conversion reactions, with $M_n$ values closer to the expected values.

Table 3. LLA polymerization in the presence of BnOH. [a]

| Ent. | Cat | [LLA]:[Cat]:[BnOH] | t (min) | conv.[b] (%) | $M_{n,\,theo}$[c] (kDa) | $M_{n,\,exp}$[d] (kDa) | Đ[d] |
|---|---|---|---|---|---|---|---|
| 1 | 1 | 200:1:2 | 1 | >99 | 14.5 | 11.8 | 1.40 |
| 2 | 1 | 400:1:2 | 2 | >99 | 28.9 | 22.9 | 1.39 |
| 3 | 1 | 800:1:2 | 4 | 99 | 57.8 | 41.7 | 1.56 |
| 4 | 2 | 200:1:2 | 1 | >99 | 14.5 | 19.4 | 1.49 |
| 5 | 2 | 400:1:2 | 2 | >99 | 28.9 | 31.3 | 1.60 |
| 6 | 2 | 800:1:2 | 6 | 65 | 37.5 | 31.4 | 1.70 |

[a] All polymerizations were performed in $CH_2Cl_2$ at room temperature, [LLA] = 2 M.
[b] Determined by $^1$H NMR spectroscopy. [c] $M_{n,\,theo}$ = ([LLA]$_0$/[BnOH]$_0$) · MW$_{,\,LLA}$ · %Conv. + $M_{w,\,BnOH}$. [d] Determined by SEC in THF, using polystyrene standards.

Regarding the molecular weights of the produced PLLAs, the ones obtained using **1** would meet the values needed for their processing and application as polymeric matrix. However, other alternatives could be considered for the polymers obtained with lower molecular weights, such as their use as biobased and biodegradable additives for plastics or other applications like drug encapsulation [35].

*Mechanistic studies and PLLA characterization*

To enhance our understanding of the LLA polymerization mechanism for these compounds, representative samples were analyzed by high-resolution matrix-assisted laser desorption/ionization (MALDI-TOF) mass spectroscopy. Given that the nature of the end-chain group depends on the polymerization mechanism, we selected polymers obtained in the presence and absence of BnOH, from Tables 1 and 3, respectively. The mass spectra displayed in Figure 2 corresponds to polymers obtained when using **1**. The spectrum on the bottom corresponds to the polymerization using **1** without BnOH, as shown, it displays a single series of peaks corresponding to the composition $\{(C_6H_8O_4) + (C_3H_4O_2)_n + H + Na\}^+$, m/z = 142.03 + *n* 72.02 + 1.01 + 22.99, from $n$ = 4 (455 g mol$^{-1}$) to the end of the record, over $n$ = 66 (4924 g mol$^{-1}$). This pattern would fit with the formation of a cyclic polymer or a linear polymer with an enolate end-group chain. In the second case, this would imply an anionic mechanism, where the initiation step involves the deprotonation of the LLA, generating an enolate, which is the responsible of the propagation step (Scheme 2). NMR spectroscopic studies confirming the enolate end group (Figure S20) and further experimental evidence are in agreement with an anionic mechanism (*vide infra*).

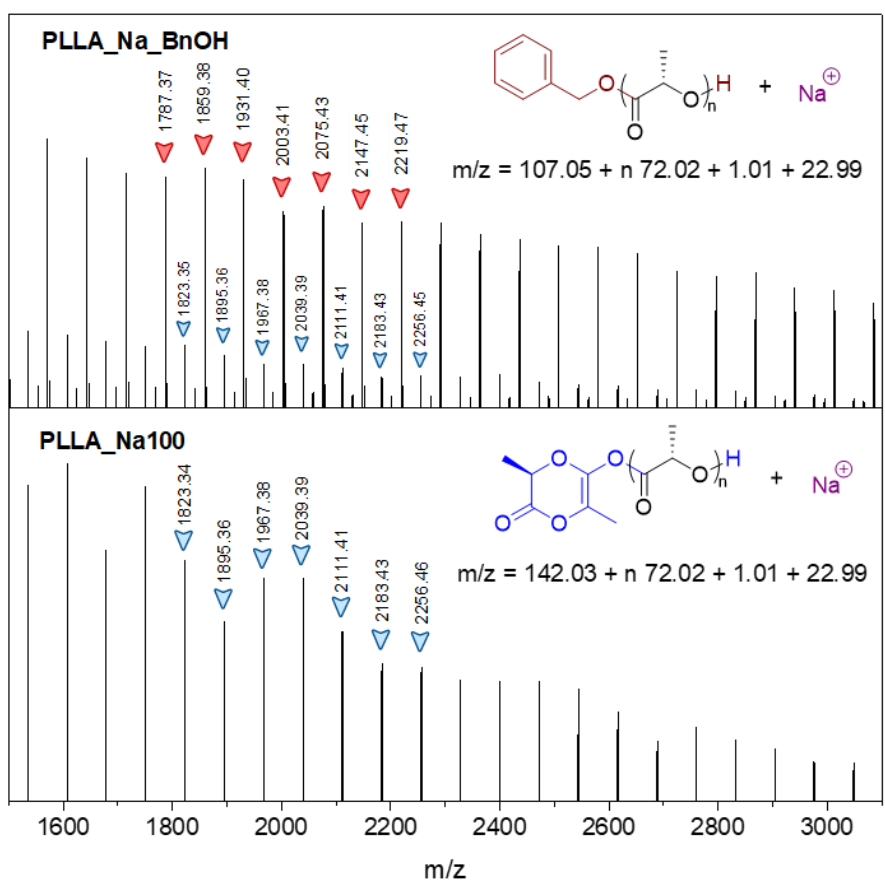

Figure 2. Mass spectrum for PLLA_Na100_BnOH (top) and PLLA_Na100 (bottom).

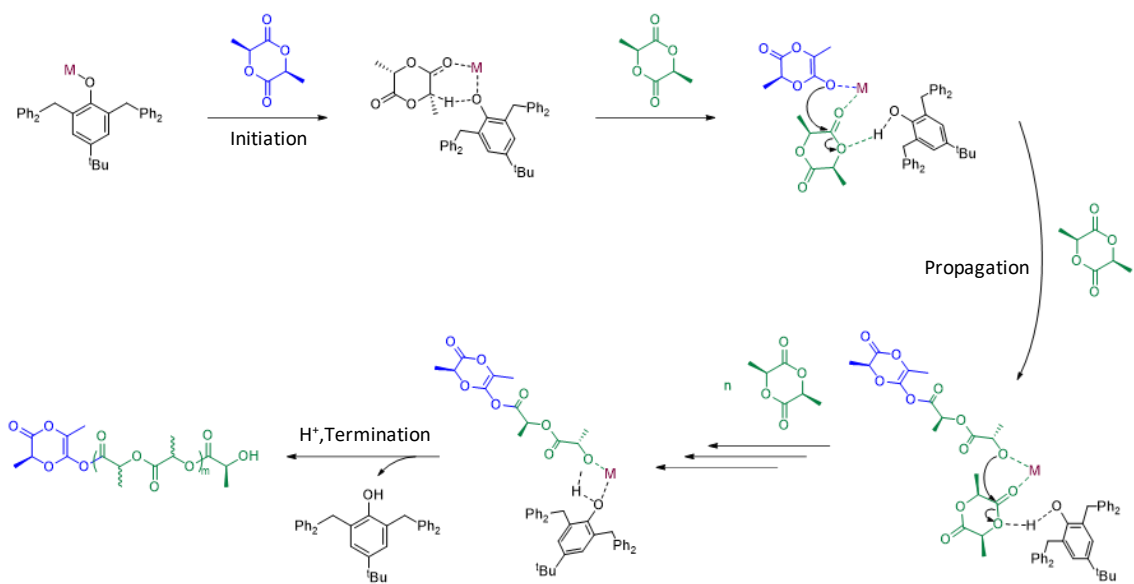

Scheme 2. Proposal of Anionic Ring-Opening Polymerization initiated by alkali-based complexes.

In the spectrum recorded for a polymer produced with the combined action of **1** and BnOH (Figure 2, top), a mayor mass series was found as $\{(C_7H_7O) + (C_3H_4O_2)_n + H + Na\}^+$, m/z = 107.05 + $n$ 72.02 + 1.01 + 22.99, starting from $n = 3$ (347 g mol$^{-1}$) to the end of the record, over $n = 67$ (4960 g mol$^{-1}$), consistent with the presence of the benzyloxide residue as the end-chain group. In addition, another minor sequence corresponding to the anionic mechanism was detected. Therefore, when BnOH was employed, although both mechanisms competed, polymerization predominantly occurred via an activated-monomer mechanism. As previously described [11, 14], in this pathway, the catalyst enhances the electrophilicity of the monomer, promoting nucleophilic attack by the alcohol. This generates the activated monomer, initiating the propagation step (Scheme 3). This mechanism was further supported by the characterization of the polymers by $^1$H NMR, as shown in Figure 3, the benzyloxide residue appears as an end-chain group. Furthermore, it was possible to calculate the molecular weight of the polymer by $^1$H NMR, that gave similar values to ones determined by GPC.

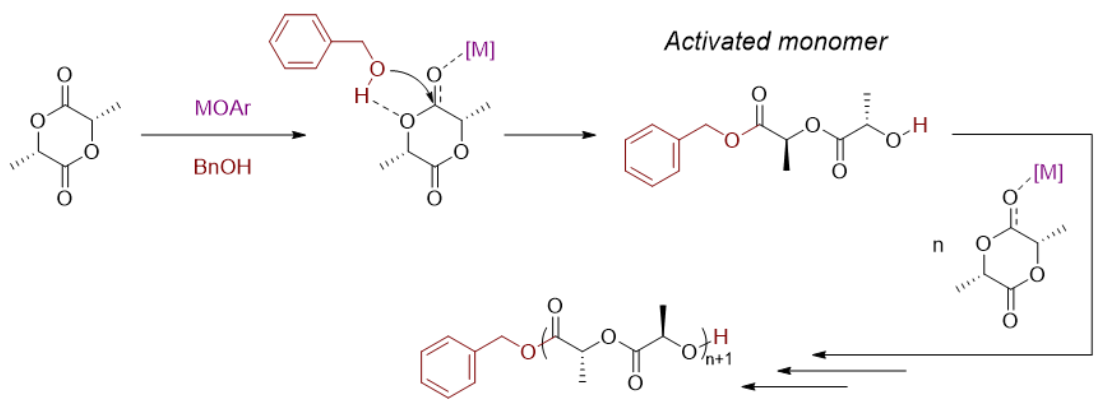

Scheme 3. Activated-monomer mechanism detected in the presence of BnOH.

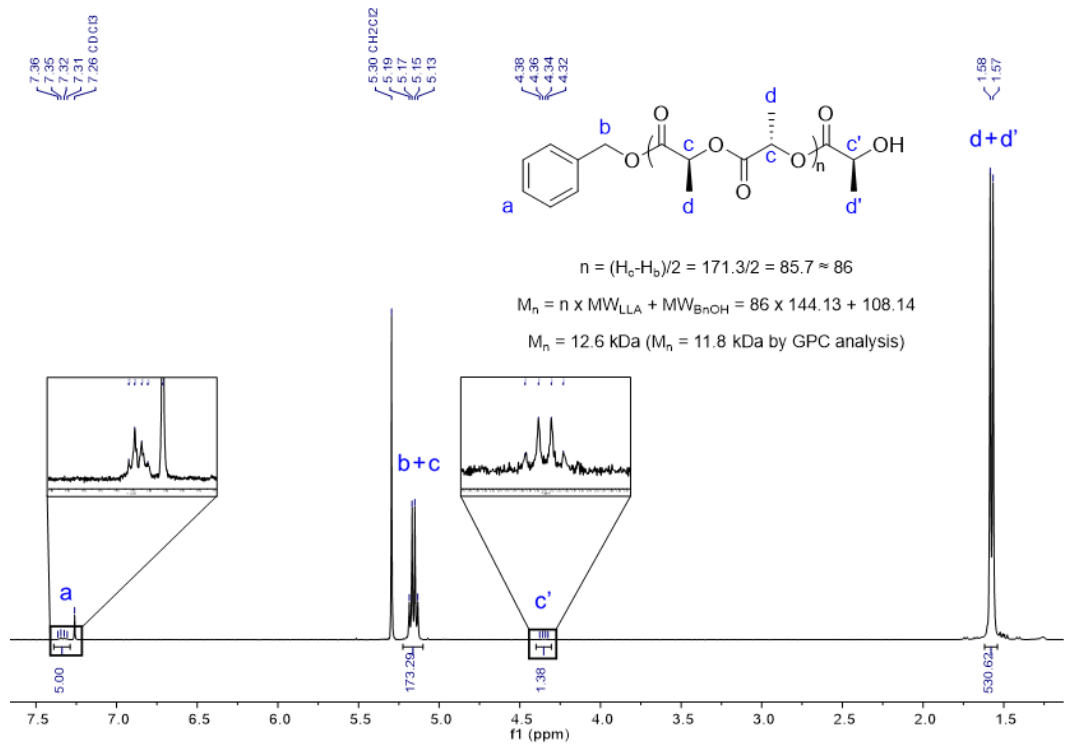

Figure 3. $^1$H NMR, CDCl$_3$: isolated polymer from the polymerization displayed in Table 3, Entry 1 ($M_n$ = 11.8 kDa, Đ = 1.40).

The stereoregularity of the polymers prepared was studied by performing the homodecoupled $^1$H NMR spectra and calculating the $P_m$ values. We observed that for the polymers obtained with BnOH as co-initiator a high isotacticity along the polymer chain is maintained, and has $P_m$ values close to 1. However, in the absence of the BnOH the anionic mechanism is prevalent, and, in these cases, $P_m$ values differ from 1 indicating the existence of epimerization processes. Considering that in PLLA only *S* stereocenters are present, if the epimerization is taking place, it could give either two *R* stereocenters next to each other or *R* stereocenters next to a *S* one, hence effectively giving *D*-lactate or *meso*-lactate units within the chain. For the calculation of the $P_m$ value, the tetrads that have been used are *rmr*, *rmm*, *mmr*, *mmm* and *mrm* (Table S1) [36], however, for some polymers a peak at 5.19 ppm reflects the presence of *meso*-lactate units (Figures S26-S29) [12]. Interestingly this effect is more evident in the PLLA obtained when compound **2** is used as catalysts, probably due to stronger basicity of the potassium aryloxide in comparison to the sodium one, which promotes deprotonation of the acidic protons.

To gain a better insight of the polymerization mechanism in the absence of BnOH, and to determine whether epimerization occurred on the chain or on the monomer before the polymerization had taken place, we studied the reaction of LLA with **1** in an equimolar ratio in a valve NMR tube and followed it by $^1$H NMR. After only 5 minutes a room temperature, full conversion to the polymer was achieved. Besides, the resonance corresponding to the protonated aryloxide ligand is clearly observed, which confirms that the formation of the initiator species implies the lactide deprotonation by the aryloxide ligand. Then the enolate generated would promote the ring opening of another lactide ring (see Scheme 2). Besides, in the NMR spectra two small quadruplets at 3.67 and 3.24 ppm are observed that can be assigned to the C-H of the end-cap groups: the resulting enolate from the lactide and the alkoxide generated (see Figures S19-S21). The signal from the alkoxide end group presents a variable integration value depending on the temperature being 1:1 for both terminal groups at -70ºC. This behavior could be due to the presence of an equilibrium where the proton from the CH next to the carbonyl group moves to the oxygen of the alkoxide end group. Interestingly, Kricheldorf [37] reported that potassium phenoxide was not able to deprotonate the lactide, however, in that case, no polymerization was observed at all, while in our system the polymerization process is evident.

As well, another interesting observation it the study of the 1:1 reaction is that as the reaction time increased, the quadruplet corresponding to PLLA became less defined, until becoming a broad signal, characteristic from an atactic PLA (Figure 4). This evidence supports that the epimerization takes place along the polymer chain, since *meso*-lactide was not observed at short reaction times. In this process, it is not clear whether it is the active growing chain or the catalyst the species responsible for the deprotonation of the CH group, which is the first step for the epimerization process [37].

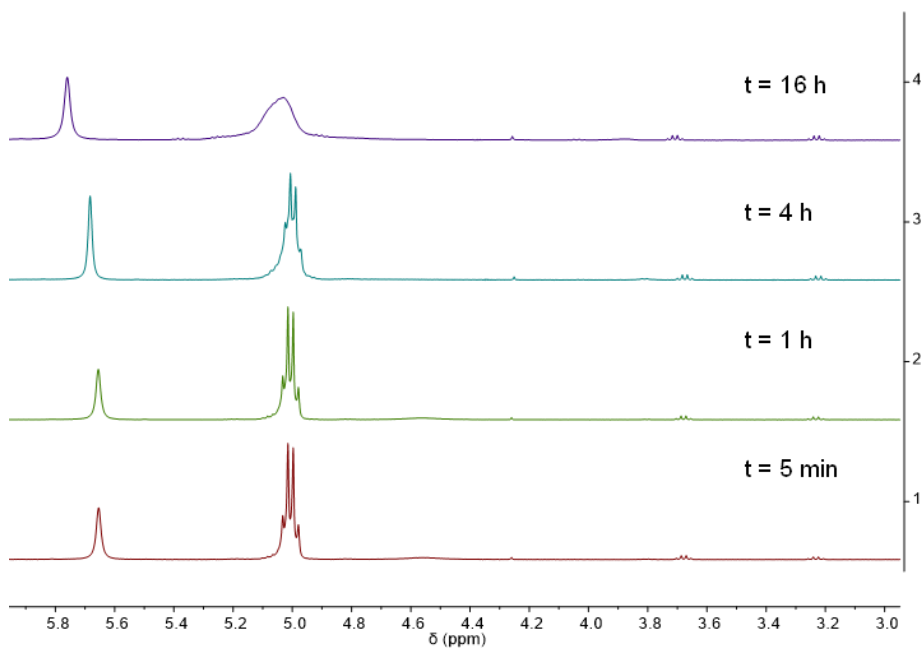

Figure 4. $^1$H NMR, $C_6D_6$: study of the reaction between LLA and **1** (ratio 1:1) over time at 25 °C.

Since the epimerization process is more favored at higher temperatures, we perform the reaction at different temperatures. The experiments displayed in Table 4 evidence that polymerization is more favorable at 25 or at 0 ºC. However, when the temperature is increased to 100 ºC, both catalysts exhibited lower activity to polymerize LLA, while the monomer epimerization is more favorable, as such, at higher temperature the formation of the *meso*-lactide in the RMN is clearly detected, being as much as 18% for the potassium derivative.

Table 4. Influence of the temperature on the ROP of LLA using alkali-based catalysts.

| Ent.[a] | Cat. | T (ºC) | Solvent | conv.[b] (%) | $S_{PLA}$[b,c] (%) | $S_{meso\text{-}LA}$[b,c] (%) |
|---|---|---|---|---|---|---|
| 1 | **1** | 0 | $CH_2Cl_2$ | 98 | >99 | 0 |
| 2 | **1** | 25 | $CH_2Cl_2$ | >99 | >99 | 0 |
| 3 | **1** | 100 | Toluene | 73 | 97 | 3 |
| 4 | **2** | 0 | $CH_2Cl_2$ | >99 | >99 | 0 |
| 5 | **2** | 25 | $CH_2Cl_2$ | >99 | >99 | 0 |
| 6 | **2** | 100 | Toluene | 33 | 82 | 18 |

[a] All polymerizations were performed employing [LLA] = 2 M, [LLA]:[Cat] = 100:1, t = 10 min.
[b] Determined by $^1H$ NMR in $CDCl_3$. [c] Selectivity: $S_x$ (%) = $Yield_x$ (%) / conv. (%)

In Scheme 4 the epimerization mechanisms are shown, whether it takes place in the chain or in the monomer in both cases it goes through an enolate intermediate. It should be noted that the fact that the epimerization takes place is a further evidence of an anionic mechanism operating. As observed, at room temperature no epimerization of the monomer is detected, while the polymerization is happening via an anionic mechanism. In this process, some stereogenic centers along the polymer chain can be deprotonated by either the active chain-end groups or the catalyst, the subsequent protonation of those carbon centers would lead to epimerization (in the figure only the catalyst is displayed as deprotonating agent, top part). At the bottom of Scheme 4, a pathway for the LLA epimerization promoted by the catalyst is shown, which involves the lactide deprotonation to generate an enolate. The formation of meso-lactide is more favored at higher temperature, and this process is competing with the polymerization which prevents the formation of higher molecular weight PLAs.

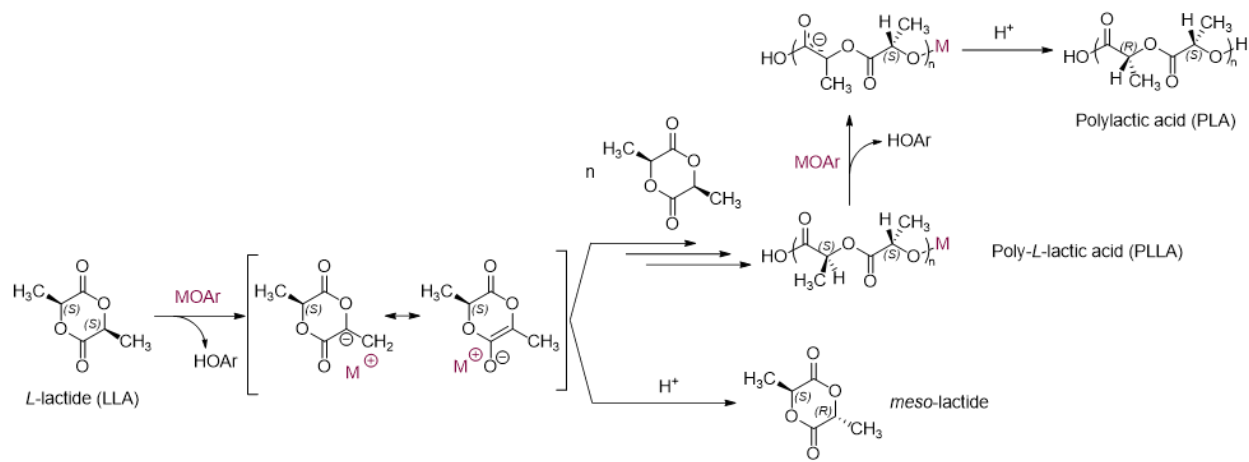

Scheme 4. Proposed epimerization mechanism in the presence of alkali-based catalysts.

To further characterize the properties of the PLLA synthetized with the different catalytic systems, and to analyze the influence of molecular weight, end-chain groups and microstructure, representative polymerization conditions were selected and scaled up to produce films and to perform their thermal and mechanical characterization. Thus, we chose samples produced using **1** and **2** with different monomer-to-catalyst ratios (1:100 and 1:200) in the absence of BnOH, as well as the closest polymers, in terms of weight average molecular weight ($M_w$), obtained in the presence of BnOH as co-initiator.

Since epimerization was detected in several samples, their specific optical rotation was measured to determine the *D*-isomer percentage presents in the polymeric chains. In accordance with the NMR spectra, the polymers produced using BnOH as co-initiator maintained stereoregularity along the polymer chain, and the resulting PLLAs is highly isotactic with low *D*-content (< 2%). Moreover, comparing the activity of both catalysts, **2** induced more epimerization than **1** resulting in values of *D*-content around 12%, confirming the trend observed by homodecoupled $^1$H NMR spectra. In Table 5, the selected samples are reported including their $M_w$, *D*-contents and their thermal properties. It is important to note that despite the different *D*-contents detected in the obtained polymers, all have been categorized as PLLA highlighting that they are all obtained from LLA.

The evaluation of their thermal transitions by Differential Scanning Calorimetry (DSC) corroborated clear differences between the samples with the benzyloxide end-chain group and those synthesized via an anionic mechanism. As shown in Figure 5, the samples obtained in the presence of BnOH displayed higher $T_m$ and $\chi_c$ compared to the other polymers. These observations are attributed to the formation of bigger crystals with a more defined morphology, which would be in agreement with the isotactic microstructure, as previously observed by homodecoupled $^1$H NMR. As shown in Table 5, the presence of the *D*-isomer clearly disrupts the crystallinity of the polymer and induces a decrease of the $T_m$, $\Delta H_m$ and $T_g$ values up to the point of hindering the crystallization of the PLLA, giving completely amorphous polymers for *D*-contents higher than 10%, as previously reported [15, 17].

Moreover, PLLA_K200 showed the lower values of $T_m$, $\Delta H_m$ and $T_g$, even though it has similar *D*-content as PLLA_K100. This behavior could be related to the presence of *meso*-lactate units predominantly, since it has been described that not only the percentage in *D*-content but also the distribution along the chain influences the thermal behaviour [15]. Thus, we could conclude that when using the potassium catalyst **1** without co-initiator, particularly for 1:200 ratio, *L*-lactate units are epimerized in *D*-lactate but also in *meso*-lactate, which has a clear effect on the PLA microstructure. The difference between the contribution of *D*- and *meso*-lactate in the disruption of the molecular chain tacticity of *L*-lactide-rich PLAs have been previously reported for *L*-lactide-rich poly(*L*-co-*D*-lactides), and poly(*L*-co-*meso*-lactides) [15].

The thermal stability of the different polymers has also been analyzed. The polymers prepared showed values close to those of commercial PLA, in the range of 290-326 °C and 228-254 °C for the maximum degradation temperature ($T_{max}$) and the initial degradation temperature ($T_{5\%}$).

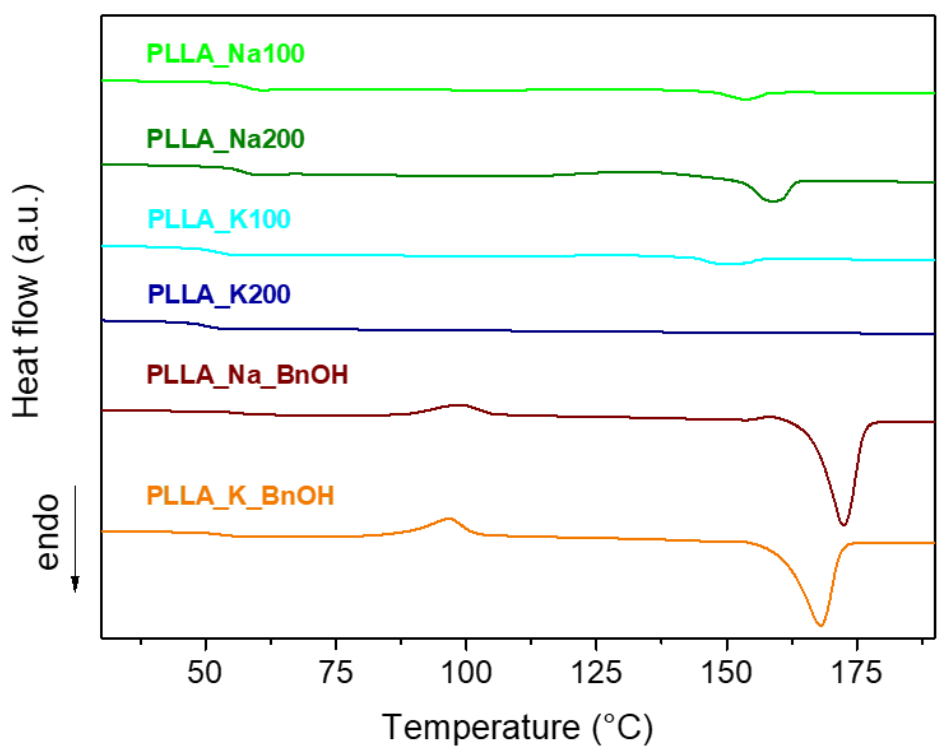

Figure 5. Compilation of DSC thermograms for the selected PLLAs (second heating).

Table 5. Compilation of the intrinsic properties of the selected PLLAs.

| Sample | $M_n$ [a] (kDa) | *D*-cont. [b] (%) | $T_g$ [c] (°C) | $T_{cc}$ [c] (°C) | $\Delta H_{cc}$ [c] (J g$^{-1}$) | $T_m$ [c] (°C) | $\Delta H_m$ [c] (J g$^{-1}$) | $X_c$ [d] (%) |
|---|---|---|---|---|---|---|---|---|
| PLLA_Na100 | 97.0 | 6.4 | 56 | 131 | 3.4 | 154 | 4.0 | 0.6 |
| PLLA_Na200 | 217.1 | 2.9 | 56 | 133 | 11.5 | 159 | 13.0 | 1.6 |
| PLLA_K100 | 93.8 | 11.3 | 52 | 129 | 3.5 | 150 | 4.2 | 0.8 |
| PLLA_K200 | 108.8 | 12.3 | 49 | - | - | 144 | 0.4 | 0.4 |
| PLLA_Na_BnOH | 65.1 | 0.7 | 57 | 99 | 10.7 | 172 | 53.5 | 46.0 |
| PLLA_K_BnOH | 53.4 | 1.6 | 53 | 97 | 15.5 | 168 | 51.7 | 38.9 |

[a] Determined by SEC in THF, using polystyrene standards. [b] Determined by specific optical rotation, see Equation (1). [c] Determined by DSC (second heating). [d] Determined by Equation (2).

Given the great differences in their isotacticity and thermal properties amongst the polymers prepared, we aimed to compare their mechanical properties. However, polymers with benzyloxide as end-chain group could not be characterized due to the impossibility of producing appropriated films, because of their lower $M_w$. Tensile tests were conducted to explore whether the *D*-content could affect the flexibility of these amorphous PLLAs. In general, the samples can be divided into two main groups. The samples produced with a 100:1 monomer:catalyst ratio showed lower Young's Modulus (E ≈ 1.8 GPa) and tensile strength ($\sigma_b$) than those obtained with a 200:1 monomer to catalyst ratio (E ≈ 2.3 GPa). Nevertheless, a clear dependence of mechanical properties with the *D*-content was not observed, probably due the effect of other factors, such as the different microstructures and molecular weights obtained. Remarkably, the mechanical properties observed were similar to those of commercially available PLA even higher values of elongation at break were registered. This behavior evidence that the epimerization reactions lead to an increase of the PLLA flexibility compared with commercial PLAs (see Table S7).

*PLLA chemical upcycling*

Different studies were conducted to assess the activity of **2** in the alcoholysis reaction of PLLA to convert it into different alkyl lactates (R-LLA, R = Bn, Me, Et) mediated by the corresponding alcohol (ROH, R = Bn, Me, Et). The initial experiments were performed in $CH_2Cl_2$ with a catalyst loading of 1 mol%, using an excess of the alcohol (see Table 6, for the calculations see Figures S36-S41) to investigate the influence of the temperature and the nucleophile used. As shown in table 6, this catalyst is active even at room temperature. Furthermore, full selectivity toward alkyl lactates was achieved in most cases, except when BnOH was employed, leading to the formation of a small amount (2-3 %) of lactic acid oligomers (OLA) as by-products.

Table 6. Influence of the temperature over the PLLA alcoholysis promoted by **2**. [a]

| Ent. | Alcohol (ROH) | T (°C) | $X_{PLLA}$ [b] (%) | $Y_{R-LLA}$ [b] (%) | $S_{R-LLA}$ [c] (%) |
|---|---|---|---|---|---|
| 1 | BnOH | 40 | >99 | 98 | 98 |
| 2 | BnOH | 25 | >99 | 97 | 97 |
| 3 | MeOH | 40 | >99 | >99 | >99 |
| 4 | MeOH | 25 | >99 | >99 | >99 |
| 5 | EtOH | 40 | >99 | >99 | >99 |
| 6 | EtOH | 25 | >99 | >99 | >99 |

[a] All reactions were performed in $CH_2Cl_2$, [LLA] = 70 mg mL$^{-1}$, $n_{ester}$:$n_{ROH}$ = 1:7.
[b] Determined by $^1$H NMR spectroscopy. [c] $S_{R-LLA}$ = $Y_{R-LLA}$ / $X_{PLLA}$.

To further optimize the process, the alcohol loading was reduced (see Table 7). In the experiments with BnOH, the amount of alcohol could be reduced from 7 equivalents to just 1, without a decrease in PLLA conversion. However, the selectivity for Bn-LLA, in favor of the OLAs, decreased with the 1:1 ratio. When MeOH was used, conversion to Me-LLA was nearly quantitative, even with a stoichiometric ratio for the ester linkages and the alcohol. Meanwhile, in the reaction of PLLA with EtOH, high conversions were achieved but reducing the alcohol excess also decreased selectivity to Et-LLA. Among these results, the use of MeOH provided the best reaction conditions, with higher conversion while using the lowest quantity of alcohol.

Table 7. Influence of the alcohol amount over the PLLA alcoholysis in the presence of **2**. [a]

| Ent. | $n_{ester}:n_{ROH}$ | Alcohol (ROH) | $X_{PLLA}$ [b] (%) | $Y_{R-LLA}$ [b] (%) | $S_{R-LLA}$ [c] (%) |
|---|---|---|---|---|---|
| 1 | 1:7 | BnOH | >99 | 99 | 99 |
| 2 | 1:2 | BnOH | >99 | 96 | 96 |
| 3 | 1:1 | BnOH | 98 | 69 | 70 |
| 4 | 1:7 | MeOH | >99 | >99 | >99 |
| 5 | 1:2 | MeOH | >99 | >99 | >99 |
| 6 | 1:1 | MeOH | >99 | 99 | 99 |
| 7 | 1:7 | EtOH | >99 | >99 | >99 |
| 8 | 1:4 | EtOH | 99 | 89 | 90 |
| 9 | 1:2 | EtOH | 93 | 77 | 83 |

[a] All reactions were performed in $CH_2Cl_2$ at room temperature, [LLA] = 70 mg mL$^{-1}$. [b] Determined by $^1$H NMR spectroscopy. [c] $S_{R-LLA} = Y_{R-LLA} / X_{PLLA}$.

Following the previous optimization, the activity of **1** was compared to its potassium analogue. Although high conversions were achieved, it is noted that this catalyst exhibits lower activity in most of the experiments (see Table 8). Again, for **1**, using MeOH leads to better conversions, followed by BnOH, and, finally, EtOH.

Table 8. PLLA alcoholysis using **1** as catalyst. [a]

| Ent. | $n_{ester}:n_{ROH}$ | Alcohol (ROH) | $X_{PLLA}$ [b] (%) | $S_{R-LLA}$ [b] (%) | $Y_{R-LLA}$ [c] (%) |
|---|---|---|---|---|---|
| 1 | 1:7 | BnOH | >99 | 98 | 98 |
| 2 | 1:2 | BnOH | 99 | 86 | 85 |

| 3 | 1:7 | MeOH | >99 | >99 | >99 |
| 4 | 1:2 | MeOH | 96 | 91 | 87 |
| 5 | 1:7 | EtOH | 99 | >99 | 99 |
| 6 | 1:2 | EtOH | 94 | 48 | 45 |

[a] All reactions were performed in $CH_2Cl_2$ at room temperature, [LLA] = 70 mg mL$^{-1}$. [b] Determined by $^1$H NMR spectroscopy. [c] $S_{R-LLA} = Y_{R-LLA} / X_{PLLA}$.

On Scheme 5 a mechanism for the alcoholysis process is shown. The presence of the catalyst is key to enhance the electrophilicity of the carbonyl group and favors the nucleophilic attack of the corresponding alcohol present, which leads to the cleavage of the chain [22]. The fact that our catalysts are so active makes this process very selective to the production of the alkyl lactate in very mild conditions.

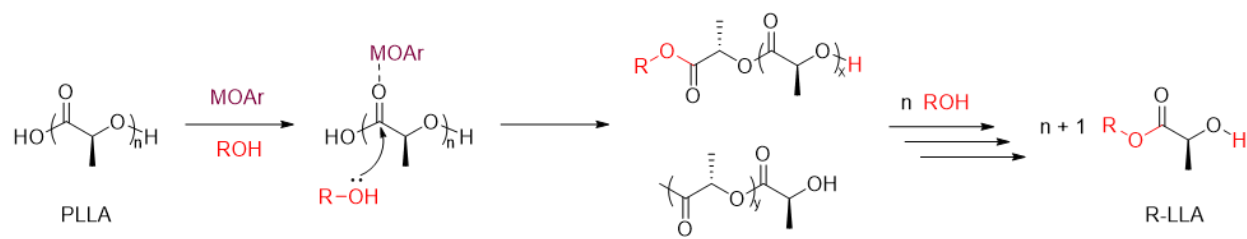

Scheme 5. Proposed mechanism for alcoholysis of PLLA in the presence of alkali-based catalysts **1** and **2**.

At this point, we selected the catalytic system involving **2** and MeOH since it was the most efficient. Furthermore, our aim was to develop a greener process by replacing $CH_2Cl_2$ with more environmentally friendly solvents. Initially, we tested tetrahydrofuran (THF) and 2-methyltetrahydrofuran (2-MeTHF), but due to their coordinating ability, the catalyst's activity was reduced, and quantitative yield could not be achieved (see Table 9). Nevertheless, we still observed high selectivity for Me-LLA. Another alternative was ethyl acetate (EtOAc), in this case, also full conversion of PLLA and high selectivity to Me-LLA were achieved, with only a very small amount of OLA as a by-product. It should be noted that the polymer solubility is lower in these solvents, so the reaction medium had to be more diluted.

Table 9. Influence of the solvent over the PLLA methanolysis using **2** as catalyst. [a]

| Ent. | Solvent | $X_{PLLA}$ [b] (%) | $S_{R-LLA}$ [b] (%) | $Y_{R-LLA}$ [c] (%) |
|---|---|---|---|---|
| 1 | $CH_2Cl_2$ | >99 | >99 | >99 |
| 2 | THF | 99 | 98 | 97 |
| 3 | 2-MeTHF | 99 | 98 | 97 |
| 4 | EtOAc | >99 | 96 | 96 |

[a] All reactions were performed at room temperature, [LLA] = 35 mg mL$^{-1}$, $n_{ester}$:$n_{ROH}$ = 1:2. [b] Determined by $^1$H NMR spectroscopy. [c] $S_{R-LLA}$ = $Y_{R-LLA}$ / $X_{PLLA}$.

Finally, a proof of concept using commercial products was carried out under the most efficient conditions. In both experiments, a cup and a teaspoon of PLA were employed for their transformation into Me-LA (Figure 6). Full conversion to Me-LA was achieved in $CH_2Cl_2$, in the presence of 1 mol% of compound **2** and 2 equivalents of MeOH per ester linkage unit.

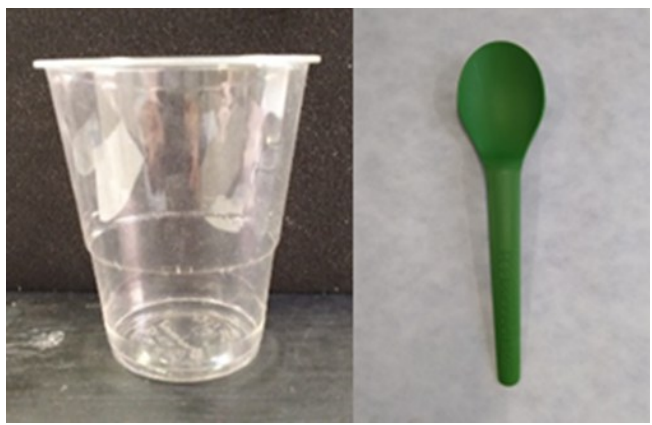

Figure 6. PLA waste employed for chemical upcycling in the presence of **1**.

Finally, in order to demonstrate the potential application of Me-LLA as a green solvent, it was used as a co-solvent for the processing of polymers by the electrospinning technique. Particularly, one of the PLLAs sinthesised previously was processed into electrospun fibers. Typically, a mixture of chloroform and dimethylformamide (DMF) in a ratio of 85/15 (w/w) is used to prepare the solution which has to be processed. Hence, we employed Me-LLA as a greener solvent to replace DMF in the mixture. Figure 7 shows that electrospun fibers of PLLA were successfully obtained using the product of the methanolysis of PLLA, corroborating the possibility of alternatively closing the loop of PLLA lifecycle by generating added-value products.

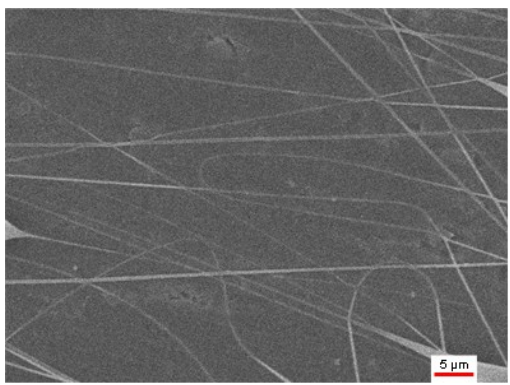

Figure 7. Electrospun fibers obtained from a PLLA solution in a mixture of $CHCl_3$/Me-LLA (85/15, w/w) observed by SEM.

**Conclusions**

The aryloxide compounds prepared based on non-toxic metals, potassium and sodium, are highly efficient catalysts. These compounds are able to polymerize under mild conditions *L*-lactide to generate PLLA with different microstructures, and thermal and mechanical properties depending on the experimental synthetical conditions.

As such, it is possible to obtain highly isotactic polymers when using a co-initiator such as BnOH, whereas in the absence of the co-initiator an anionic mechanism is prevalent and leads to high molecular weight polymers, with similar values to commercial PLLA. In this case, epimerization processes of the *L*-lactide takes place, as detected in the NMR data, that leads to the presence of a percentage of *D*-isomer content which has a clear impact on the thermal properties of the polymers. Even a small amount of *D*-lactate units affect the crystallinity and the thermal properties of the polymer. Furthermore, the study of the mechanical properties evidenced that the presence of *D*-lactate increases the flexibility of the polymer in comparison to the commercial PLLAs.

Furthermore, to complete the loop we have also studied the upcycling reactions using the same compounds as catalysts and again these alkali metals aryloxide compounds are highly active catalysts for the PLLA alcoholysis and are able to perform the PLLA chemical upcycling within 15 minutes at room temperature. The best results are achieved using MeOH and it is possible to transform the PLLA into Me-LLA even when a green solvent such as ethyl acetate was used.

SUPPORTING INFORMATION

The Supporting Information is available free of charge at www.acs.org

- Synthesis of alkali-metal based complexes, general procedure of ROP of L-lactide, general procedure of PLLA upcycling, NMR Spectra, compounds characterization, polymerization monitoring and polymers characterization, kinetics studies of the LLA polymerization, Homodecoupled $^1$H NMR spectra, chemical upcycling studies, Single-Crystal X-Ray Structure Determination, GPC traces, Mass Spectra, TGA thermograms, DSC thermograms, Tensile Tests curves and values.

AUTHOR INFORMATION

Corresponding Authors

**Marta E. G. Mosquera** - Departamento de Química Orgánica y Química Inorgánica, Instituto de Investigación en Química "Andrés M. del Río" (IQAR), Universidad de Alcalá, Campus Universitario, Alcalá de Henares, Madrid 28871, Spain. https://orcid.org/0000-0003-2248-3050 Email: martaeg.mosquera@uah.es.

**Valentina Sessini** - Departamento de Química Orgánica y Química Inorgánica, Instituto de Investigación en Química "Andrés M. del Río" (IQAR), Universidad de Alcalá, Campus Universitario, Alcalá de Henares, Madrid 28871, Spain. https://orcid.org/0000-0003-1205-4586 Email: valentina.sessini@uah.es.

AUTHOR CONTRIBUTIONS

The manuscript was written through contributions of all authors. All authors have given approval to the final version of the manuscript. Conceptualization, methodology and supervision, M.E.G.M.[†] and V.S.[†]; formal analysis and investigation, C.R.[†], M.P.[†§] and L.G.[†‡]; project management, M.E.G.M[†]; writing-original draft preparation, M.E.G.M.[†], V.S.[†] and C.R.[†]; writing-review and editing, M.E.G.M.[†] and V.S.[†] All authors have read and agreed to the published version of the manuscript.

ACKNOWLEDGMENT

   This research was supported by Ministerio de Ciencia e Innovación (Spain) (PID2021-122708OB-C31 AEI/10.13039/501100011033), Comunidad de Madrid (EPU-INV/2020/001), and the University of Alcalá (UAH-AE-2017-2). V.S. would like to thank the Ministerio de Ciencia e Innovación (Spain) and the European Community (grant number: RYC2021-033921-I) for the financial support. C.R would like to thank University of Alcalá for a fellowship. Finally, the authors acknowledge the support provided by COST Action CA20101 Plastics monitoRIng detectiOn RemedIaTion recovery—PRIORITY, supported by COST (European Cooperation in Science and Technology, www.cost.eu, accessed on 22 April 2022). The authors gratefully acknowledge Dr. Adrián Leonés and Dr. David Sánchez-Roa for their assistance in recording homodecoupled $^1$H NMR and determining the specific optical rotation of the polymers.

35. (a) Lee, B. K.; Yun, Y.; Park, K., PLA micro- and nano-particles. *Adv Drug Deliv Rev* **2016**, *107*, 176-191. (b) Sahini, M. G., Polylactic acid (PLA)-based materials: a review on the synthesis and drug delivery applications. *Emerg Mater* **2023**, *6*, 1461-1479.

36. Singha Roy, S.; Sarkar, S.; Chakraborty, D., Determination of Polylactide Microstructure by Homonuclear Decoupled (1) H NMR Spectroscopy. *Chem Rec* **2021**, *21*, 1968-1984.

37. Kricheldorf, H.R.; Kreiser-Saunders, I. Polylactones, 19. Anionic polymerization of L-lactide in solution. *Makromol Chem* **1990**, *191*, 1057–1066.
40